\documentstyle[seceq,wrapfig]{ptptex}

\def\abs#1{\left| #1\right|}
\def\ltap{\ \raise.3ex\hbox{$<$\kern-.75em\lower1ex\hbox{$\sim$}}\ }
\def\gtap{\ \raise.3ex\hbox{$>$\kern-.75em\lower1ex\hbox{$\sim$}}\ }

\notypesetlogo  

\preprintnumber{
 UW-PTH/95-17,hep-ph/9511218}

\title{
New~Issues~In~Low~Energy
Dynamical~Supersymmetry~Breaking
}

\author{
Ann E. {\sc Nelson}\footnote{  E-mail address: anelson@phys.washington.edu}
}

\inst{
Physics Department,  Box 351560, University of Washington,
Seattle,~WA~98195-1560,~USA
 }

\recdate{
October 1995
}

\abst{ Comparatively
simple models, with   low energy structure similar to that of the MSSM,
but with far fewer arbitrary parameters, can be constructed in which
supersymmetry is dynamically broken at low energies. The phenomenology
of these models is somewhat different than that of the usual scenario
with supersymmetry broken in a hidden sector.
}

\begin{document}

\maketitle

\section{Introduction}
I would like to report on some work done recently with Michael Dine, Yossi Nir
and Yuri Shirman on models in which  supersymmetry breaking occurs
dynamically and is communicated to the  squarks, sleptons and gauginos via
renormalizable gauge interactions\cite{rf:new}. Our motivations for
constructing
such models were threefold.
\begin{enumerate}
\item Dynamical Supersymmetry Breaking (DSB) has the potential to explain the
hierarchy between the scales of weak and gravitational
interactions\cite{rf:witten}.
\item If supersymmetry breaking is communicated to the squark and sleptons via
gauge interactions, sufficient degeneracy to be compatible with Flavor Changing
Neutral Current Constraints (FCNC) is automatic.
\item Since we do not know either the mechanism or the scale of supersymmetry
breaking, we should explore as many possibilities as we can.
\end{enumerate}
\smallskip
For the technicolor enthusiasts in the audience, let me add a couple of other
motivations for exploring models of low energy DSB.
\begin{enumerate}
\item Interesting nonperturbative effects occur in these models, which are
under
theoretical control.
\item One might try and also explain dynamically in a DSB model   puzzling
features of both the standard model and the MSSM such as the quark and lepton
mass spectrum and the number of families. Such a model would have the same
theoretical attraction as technicolor. Supersymmetry has the advantage of
both enlarging and constraining the dynamical possibilities. While in
technicolor
models it is hard to see how to eliminate unwanted exact chiral symmetries,
in supersymmetric theories there are naturally light scalars and Yukawa
couplings.
\end{enumerate}\smallskip
Our model building strategy to date has been simple but not
elegant--our models have three ``sectors''. The superpotential does not mix the
different sectors--only gauge interactions connect them.  In the heaviest
sector,
called the ``supercolor'' sector, supersymmetry is dynamically broken via gauge
interactions at a scale of
$\sim 10^6-10^7$ GeV. We have recently discovered a large number of new
possibilities for this sector \cite{rf:new}, and there are clearly a huge
number
of still unexplored possibilities. To communicate supersymmetry breaking, we
gauge
a global symmetry of the supercolor sector (the ``messenger'' gauge
group).
We then
add  a few superfields carrying the messenger symmetry and having
  superpotential
couplings with particles carrying ordinary color and electroweak gauge
interactions (the messenger sector). These ordinary gauge interactions feed
down
supersymmetry breaking radiatively into the ordinary sector.

The resulting models are somewhat different from the usual
Minimal Supersymmetric Standard Model (MSSM); the major differences are
summarized in the following table.
\begin{center}
\let\tabularsize\normalsize
\begin{tabular}{llp{.7\textwidth}}
&&\\
&{\bf Low Energy DSB} &{\bf MSSM }\\&&\\
{\it Supersymmetry} &Spontaneous, nonperturbative&
``Soft'' (assumed to occur in \\
{\it Breaking}&$M_S\sim M_p e^{-4 \pi^2 / g^2 }$ & ``hidden'' sector)   at
$\sim
10^{11}$ GeV\\ &&\\
{\it Messenger of}&gauge interactions at&gravitational, Planck scale\\
{\it SUSY Breaking}&$10^4-10^6$ GeV& interactions at $10^{19}$ GeV\\&&\\
{\it Lightest}&gravitino&mixture of photino,\\
{\it Supersymmetric}&$m_{3/2}\sim 1$ keV &  Zino, higgsino \\
{\it Particle}&&\\ &&\\
{\it Superpartner}&Calculable in terms of &$\sim 100$ free ``soft''
parameters\\
{\it Masses}&2 parameters  arising in&reducable to $\sim 4$ parameters\\
&renormalizable Lagrangian&via theoretical assumptions\\&&\\
{\it FCNC}&small due to accidental&assumed to be small, requires\\
&approximate flavor symmetry& squark, slepton degeneracy\\&&\\
{\it Cosmology}&gravitino ``hot'' dark matter&LSP cold dark matter, gravitino\\
&possible TeV mass cold & decays a problem for nucleosynthesis,\\
&dark matter& gravitationally coupled light\\
&&scalars a serious problem\\&&\\
\end{tabular}
\end{center}
\section{The Dynamical Supersymmetry Breaking Sector}
There is a simple criterion for models which exhibit
dynamical supersymmetry breaking \cite{rf:ads}. If a theory has no
flat directions, and it has a global symmetry which is spontaneously
broken, then supersymmetry is spontaneously broken.
Here I describe a new   model  which
satisfies this criterion and which can serve as a simple supercolor sector.

The gauge group is $SU(6) \times U(1)\times U(1)_m$.
(The $U(1)_m$ symmetry plays the role of the ``messenger'' gauge group.
This theory has a stable supersymmetry breaking ground state  whether
or not the
$U(1)_m$ is gauged.) The chiral superfields are:
\begin{equation}A_{+2,0}\ \ \ F_{-5,0}\ \ \ \bar F^{\pm}_{-1,\pm 1}\ \ \
\bar F^0_{-1,0}\ \ \ S^\pm_{+6,\pm1}\ \ \ S^0_{+6,0},
\end{equation}
($A=15$, $F=6$, $\bar F=\bar 6$ and $S=1$ of $SU(6)$, the subscripts
give the $U(1)$ charges.)
For the superpotential we take
\begin{equation}
W=\lambda A\bar F^+\bar F^-+\gamma F(\bar F^+S^-+\bar
F^-S^+)+\eta F\bar F^0 S^0.\end{equation}
Here we have imposed a discrete symmetry,
\begin{subequations}\begin{eqnarray}
A\rightarrow A,&\ \ \ F\rightarrow +iF,\\
\bar F^\pm\rightarrow -i\bar F^\mp,&\ \ \
\bar F^0\rightarrow -i\bar F^0,\\
S^\pm\rightarrow S^\mp,&\ \ \ S^0\rightarrow
S^0, \end{eqnarray}\end{subequations} under which  the
$U(1)_m$ gauge fields change sign. This discrete symmetry guarantees that
  a $D$ term for $U(1)_m$ will not be generated, which is good because in
earlier
model building attempts with a $U(1)$ messenger group \cite{rf:dns} the
generation of a D term required that the messenger sector be complicated and
slightly fine-tuned to avoid undesirable symmetry breaking patterns.
This theory also has a nonanomalous, global $U(1)_R$ symmetry. By looking at
the
effective theory along various directions in field space, it is possible to
show
that gaugino condensation in an unbroken $SU(2)$ subgroup of the
$SU(6)$ gives   a term in the effective superpotential which
leads to spontaneous breaking of this R symmetry and of supersymmetry. The
supersymmetry breaking minimum may be systematically studied in the limit where
the superpotential couplings are weak, and we find that
  the vev's of the fields   have the following form:
\begin{subequations}\begin{eqnarray} A=&
\pmatrix{\sqrt{{v^2\over2}+a^2}\sigma_2&&\cr &a\sigma_2&\cr &&a\sigma_2\cr},
\ \ S^0=c,\\
\bar F^-=&\pmatrix{v\cr 0\cr 0\cr 0\cr0\cr0\cr},\ \
\bar F^+=\pmatrix{0\cr v\cr 0\cr 0\cr0\cr0\cr},\ \
\bar F^0=\pmatrix{0\cr 0\cr b\cr 0\cr0\cr0\cr},\ \
F^{0}=\pmatrix{0\cr0\cr b\cr 0\cr0\cr0\cr},  \end{eqnarray}\end{subequations}
and all other vev's
vanish.
A linear combination of messenger hypercharge and a subgroup of the $SU(6)$  is
unbroken in this ground state.

\section{The Messenger Sector}
Although it is nice to have an explicit model of the supersymmetry breaking
sector, for the foreseeable future we are unlikely to have direct experimental
access to this physics, but at best expect to probe the spectrum of
superpartners
of the already observed particles. In our model building approach,
supersymmetry breaking is transmitted to the ordinary superpartners via an
intermediate sector, the ``messenger''. The ordinary superspectrum then is
quite
insensitive to the details of the supercolor sector, but does crucially depend
on
the choice of messenger sector. Fortunately, we find that an extremely simple
choice here will lead to acceptable superpartner masses.

In addition to the chiral superfields of the previous section, we include
a gauge singlet
$X$, two fields
$\phi^+$ and $\phi^-$ with messenger charge $\pm 1$, and   vector-like
quark and lepton fields, $q$, $\bar q$, $\ell$ and $\bar \ell$ carrying
ordinary
$SU(3)\times SU(2)\times U(1)$. For this set of fields we take the
superpotential
to be
\begin{equation}
W_X= k_1\phi^+\phi^-  X+
{1 \over 3}\lambda X^3 + k_3 X \bar \ell \ell + k_4X \bar q q\ .
\end{equation}
At two loops, the scalar components of $\phi^+$ and $\phi^-$ gain a mass
squared which is  negative for a range of parameters:
\begin{equation}
m_{\phi}^2=-{1\over 2}
\left({\alpha_m\over\pi}\right)^2
m_{\chi}^2 \ln(\Lambda_6^2/m_{\chi}^2).\end{equation}
Here $\Lambda_6$ is the scale of the $SU(6)$ theory; it is roughly the
scale where the $\chi$ mass is determined.

As a result, the effective potential for $\phi^\pm$ and $X$ has the form,
ignoring for a moment the terms involving $q,\ell$ and $\bar q,\bar \ell$,
\begin{equation}m_{\phi}^2 \left(\abs{\phi^+}^2+\abs{\phi^-}^2\right) +
\abs{k_1 X \phi^+}^2 +\abs{ k_1 X \phi^-}^2
+\abs{k_1\phi^+ \phi^- +\lambda X^2}^2\ .\end{equation}
At the minimum of this potential, $\phi^+$,  $\phi^-$,
$X$ and $F_X$  have non-zero vev's.  For sufficiently
small $\lambda$, this point is a minimum with zero vev's
for the fields $q$, $\bar q$, $\ell$ and $\bar \ell$. Note that had there
been a Fayet-Iliopoulos term at one loop for $U(1)_m$, $F_X$ would
have been zero.

With nonzero $F_X$ and $\langle X \rangle$, the fields $q$, $\bar q$, $\ell$
and
$\bar \ell$ all obtain mass and have a nonsupersymmetric mass spectrum, with
some
scalars lighter than their fermionic superpartners and some heavier.
\section{Supersymmetry Breaking in the MSSM Sector}
We can now integrate out the supercolor and messenger sectors, and consider
  the
masses of squarks, sleptons and gauginos.  Loop corrections to these masses
arise when we integrate out the fields $q$, $\bar q$, $\ell$ and $\bar \ell$.
 At
one loop, for small
$\lambda$, we obtain (Majorana) masses for the
$SU(3)$, $SU(2)$ and $U(1)$ gauginos to lowest order in $F_X$:
\begin{equation}m_{\lambda_i}=c_i\ {\alpha_i\over4\pi}\ \Lambda\
,\end{equation}
where $c_1={5\over3}$, $c_2=c_3=1$, and the parameter $\Lambda$,
\begin{equation}\Lambda={F_X\over X},\end{equation}
sets the scale for {\it all} of the soft breakings in the low energy
theory. Masses squared for the squarks and sleptons appear due to gauge
interactions at two loops.  They are given by
\begin{equation}\tilde m^2 ={2 \Lambda^2} \left[
C_3\left({\alpha_3 \over 4 \pi}\right)^2
+C_2\left({\alpha_2\over 4 \pi}\right)^2
+{5 \over 3}
{\left(Y\over2\right)^2}\left({\alpha_1\over 4
\pi}\right)^2\right].\label{eq:squarks}\end{equation}
Here
$C_3 = 4/3$ for color triplets and zero for singlets; $C_2= 3/4$ for weak
doublets
and zero for singlets, and $Y$ is the ordinary hypercharge. There is no general
reason for these masses squared to be positive, and finding positive
squark masses squared is a nontrivial constraint on the allowed
messenger sector.  Fortunately they  turn out
positive for this  simple messenger sector.

Note the structure of the theory at this level.  Squarks are the most
massive scalar fields, by roughly a factor of three compared to slepton
and Higgs doublets.  Slepton singlets are the lightest scalar fields,
by still another factor of order three.  Gluinos have masses comparable
to squarks, while the Majorana component of the wino mass matrix
is comparable to that of the doublets.  Note also that the strict
degeneracy of squarks and of sleptons of the same gauge quantum numbers
is only broken by effects of order quark or lepton Yukawa couplings.
We will see that experimental constraints give masses for squarks and
gluinos in the $200-300$ GeV range. This means that $\Lambda \sim 10
\quad TeV$. This is the scale of messenger physics. The scale of the hidden
sector $SU(6)\times U(1)$ physics is larger by a factor of order
${(4 \pi)^2\sqrt\lambda\over \alpha_m k_1^2}$, about $10^3$ TeV for
coupling constants of order one.

\section{Ordinary $SU(2) \times U(1)$ Breaking}
 Note that   in this framework, it seems more difficult than in the MSSM to
explain the size of the supersymmetric term
  $\mu H_U H_D$ term in the superpotential with $\mu $ of order the weak scale.
This difficulty previously led us to consider a low energy sector more
complicated than the MSSM \cite{rf:dns,rf:dn}.
To simplify the low energy sector, we instead can use a mechanism suggested by
Leurer {\it et al.}  \cite{rf:nirseiberga}. In this mechanism, in addition to
the
usual MSSM fields, there is another singlet, $S$, with only nonrenormalizable
couplings. In particular, consider terms in the effective lagrangian of the
form:
\begin{equation}{1 \over M_p^2}\int d^4 \theta X^{\dagger} X S^{\dagger} S
+\int d^2 \theta
\left({1 \over M_p^p}XS^{2+p}+ {1 \over M_p^m}S^{m+3}+
{1\over M_p^n} S^{n+1}H_UH_D\right)\ .
\end{equation}
The first and second terms can contribute effective negative curvature
terms to the $S$ potential, and the field $S$ obtains a large vev, leading to
an effective $\mu$ term.   For example, if
$p=2$,
$m=2$ and
$n=1$, then the $\mu$ term is of order $\sqrt{F_X}$
times powers of coupling constants.

We can now analyze the Higgs potential.
 First, note that a coupling in the superpotential:
\begin{equation}W_{XH}=\lambda^{\prime} X H_U H_D\end{equation}
leads to a soft-breaking term $m_{12}^2 H_U H_D$ in the higgs{\it \
 potential}.
Here $\lambda^\prime$ must be rather small, since these masses should
be roughly of order $(\alpha_2/ \pi)^2$. This smallness is natural,
in the sense of 't Hooft, in that it can arise due to approximate
discrete or continuous symmetries. Note that the corresponding
contribution to the $\mu$ term, however, is {\it extremely} small,
far too small to be of phenomenological significance. Positive soft
supersymmetry
breaking mass squared terms for  $H_U$ and $H_D$ are generated by the same type
graphs giving slepton masses. Finally, a negative contribution to the mass
squared
for
$H_U$ arises from loops with top squarks. This contribution, although of three
loop
order, is somewhat larger than the two loop contributions because it is
proportional to the top squark mass squared.  We obtain
\begin{equation}m_{H_U}^2-m_{H_D}^2 = -{ 3\over 4 \pi^2}
y_t^2 \tilde m_t^2 \vert
\ln\left({\alpha_3 \over \pi}\right)\vert,\end{equation}
where $y_t={m_t\over v_2}$
and, from eqn. \ref{eq:squarks},
\begin{equation}m_{H_D}^2 \approx {3\over2}\left({\alpha_2\over4\pi}\right)^2
\ \Lambda^2.\end{equation}
The argument of the logarithm is the ratio of the
high energy scale, roughly of order $\Lambda$, to the stop mass.

To summarize, at energies well below the scale $\Lambda$,
the theory looks like the usual MSSM, but with well-defined
predictions for the soft breaking terms. Indeed, the Higgs potential
and all
of the soft breakings among the light states are determined
in terms of three parameters:  $\Lambda$, $\mu$, and $m_{12}^2$
(we view the $t$ quark mass as known.)  Other supersymmetry breaking terms,
such as trilinear scalar couplings, are also generated but are small.

\section{Summary}
\noindent A viable DSB model can be  found with\smallskip
\begin{enumerate}
\item no hierarchy problem--all mass scales below $M_P$ arise via dimensionless
transmutation
\item unifiable ordinary $SU(3)\times SU(2)\times U(1)$
\item  additional ``supercolor'' gauge group, {\it e.g.}
$SU(6)\times U(1)$, and additional ``messenger'' gauge group, {\it e.g.}
$U(1)_m$
\item a supercolor sector with new chiral superfields carrying
supercolor and
messenger interactions
\item a simple  messenger sector with particles in vector-like representation
of messenger group, a gauge singlet, and new vector-like quarks and leptons.
\item  radiative squark,   slepton and gaugino  masses
  proportional to their gauge couplings squared
\item ordinary superpartner spectrum calculable in terms of two unknown
parameters
\item light right handed charged sleptons likely to be found at LEP II
\item no FCNC problem
\item no dangerous pseudo Goldstone bosons
\item light (keV) gravitino
\item no cosmological difficulties.
\end{enumerate}\smallskip
This is an interesting, viable alternative to the MSSM which is worth serious
consideration. However it is in several respects theoretically unsatisfying.
The  messenger group  and the division into three different sectors seem
contrived. Perhaps the messenger sector can be eliminated by using Seiberg's
idea
that some of the ordinary quarks, leptons and gauge bosons are dual to other
degrees of freedom
\cite{rf:seiberg}, by arranging for quarks and leptons to be light composites,
or by finding new DSB mechanisms and models. We might hope to discover a model
 in which the
  supersymmetry breaking mechanism involves
particles carrying ordinary gauge quantum numbers, perhaps even the ordinary
quarks
and leptons. In such a model, the scale of supersymmetry breaking might be as
low
as $\sim  4 \pi M_{W}/\alpha_{\rm wk}$, implying a gravitino mass of $\sim 0.1
$
eV. The next to lightest supersymmetric particle, {\it i.e.} the lightest
neutralino, could then decay into a
gravitino and photon with a lifetime  of $\sim 10^{-15}$ sec. This lifetime is
short enough so that the decay could occur inside particle physics detectors,
providing a ``smoking gun'' signature that the fundamental scale of
supersymmetry
breaking is low.
\section{Other Directions to Explore?}
There are many other possibilities, both for dynamical supersymmetry breaking
and for the supersymmetry breaking messenger. One possibility is that the
``messenger'' sector could be identified with part of the GUT sector. This
would
imply that the scale of susy breaking is about $10^9$ GeV. The resulting low
energy model would ressemble the MSSM in many respects, but the usual MSSM GUT
predictions for relations among the soft masses could be modified, since the
messenger sector is not blind to GUT symmetry breaking.
Another possibility is to consider models where nonrenormalizable terms in the
superpotential are required for a stable supersymmetry breaking minimum. A
simple
class of such models has gauge group $SU(N)\times U(1)$, with chiral
superfields
\begin{subequations}\begin{eqnarray}&A\sim\left({N(N-1)\over 2},N-4\right),
\ \ \ \bar F_i\sim\left(\bar N,2-N\right), i=1\ldots N-4,\\
& S_a\sim\left(1,N\right),a=1\ldots {(N-3)(N-4)\over 2} ,
\end{eqnarray}\end{subequations}
where the representation of the gauge group is in parentheses.
The tree superpotential is\begin{equation}
W_{\rm tree}={\lambda^{ija}\over m_P} A\bar F_i\bar F_j S_a\end{equation}
while a superpotential
\begin{equation}
W_{\rm dyn}={\Lambda_N^{3+2 N\over 3}\over
\left(A^{N-2}\bar F^{N-4}\right)^{1\over 3}}\end{equation} is dynamically
generated\cite{rf:ads}. An interesting hierarchy is then generated between the
scale  of supersymmetry breaking $M_s$ and $\Lambda_N$, the scale of $SU(N)$
dynamics, with
\begin{equation}M_s\sim M_P^{3-2N\over 12+4 N}\Lambda_N^{9+6 N\over 12+4
N}\ .\end{equation}  The scale $x$ of the expectation values of the superfields
is
\begin{equation} x\sim\Lambda^{3+2N\over 6+2N}M_P^{3\over6+2N}\end{equation}so
\begin{equation}M_P\gg x\gg \Lambda_N\gg M_s \ .\end{equation} Still
more scales could be generated in theories with both renormalizable and
nonrenormalizable terms in their superpotentials. It seems worth exploring
whether
some or all of the various widely separated scales appearing in our current
ideas
about particle physics, (such as the GUT scale, the various quark  and lepton
masses, the intermediate scale associated with neutrino masses etc.), could be
generated in such a model.

\section*{Acknowledgements}
I would like to thank the organizers for inviting me to this most
interesting conference.

\end{document}